\documentclass{kapproc}
\kluwerbib

\usepackage{graphics}
\begin{document}


\articletitle[Scalable Architecture for Adiabatic Quantum
Computing of NP-Hard Problems]{Scalable Architecture for\\
Adiabatic Quantum Computing \\of NP-Hard Problems}

\chaptitlerunninghead{Scalable Architecture for Adiabatic Quantum
Computing}

\author{William M. Kaminsky* and Seth Lloyd}
\affil{Massachusetts Institute of Technology, Cambridge, MA 02139
USA}
\email{*wmk@mit.edu}





\begin{abstract}
We present a comprehensive review of past research into adiabatic
quantum computation and then propose a scalable architecture for
an adiabatic quantum computer that can treat NP-hard problems
without requiring local coherent operations. Instead, computation
can be performed entirely by adiabatically varying a magnetic
field applied to all the qubits simultaneously. Local (incoherent)
operations are needed only for: (1) switching on or off certain
pairwise, nearest-neighbor inductive couplings in order to set the
problem to be solved and (2) measuring some subset of the qubits
in order to obtain the answer to the problem.
\end{abstract}



\section{Introduction}
Adiabatic quantum computation\cite{Farhi} is a recently proposed,
general approach to solving NP-hard combinatorial minimization
problems.\cite{Garey}  It consists of constructing a set of qubits
with a time-dependent Hamiltonian $\hat{H}(t)$ whose starting
point $\hat{H}_s$ has a ground state that is quickly reachable
simply by cooling and whose final point $\hat{H}_p$ has couplings
that encode the cost scheme of a desired minimization problem. The
name ``adiabatic'' comes from the fact that if the qubits are
initialized in the ground state of $\hat{H}_s$ and if $\hat{H}(t)$
is varied slowly enough, then the qubits will overwhelmingly be in
the ground state of $\hat{H}(t)$ at all times $t$, thus in
principle completely bypassing the usual concern about local
minima in $\hat{H}_p$ confounding the search for the problem's
solution.

Sections 2 and 3 review the literature to date on adiabatic
quantum computation.  Section 4 presents the main result of this
paper: a scalable architecture for an adiabatic quantum computer.
Section 5 concludes.

\section{Literature Review: The Time Complexity of Adiabatic Quantum Computation}
The general time-complexity of adiabatic quantum computation is
still an open problem.   Intuitively, the key to fast, successful
adiabatic computation is to ensure that $\hat{H}(t)$ always
possesses a sizable spectral gap between its instantaneous ground
and excited states, thus allowing one to vary $\hat{H}(t)$ quickly
while still not ever providing enough energy to excite the system
significantly out of its instantaneous ground state. Determining
the asymptotic time complexity of adiabatic quantum computation
remains an open problem because for $n$ qubits, $\hat{H}(t)$ is a
$2^n \times 2^n$ matrix, and hence analytically obtaining lower
bounds on the spectral gap is extremely difficult unless
$\hat{H}(t)$ possesses significant symmetry.

Presently, there are three major results on the time complexity of
adiabatic quantum computation.   Together, they yield the general
picture that for essentially unstructured NP-hard minimization
problems like finding the ground state of a random-field magnet,
adiabatic quantum computing should offer polynomial
speedups---probably in the range of cube-root to sixth-root---over
a Metropolis algorithm (simple cooling) approach to the same
problems.  Furthermore, for structured NP-hard minimization
problems such as 3-SAT and MAX CLIQUE, adiabatic quantum
computation can quite possibly offer exponential speedups over any
classical algorithm (assuming $P\not=NP$)---at least on average.

More specifically, the first major time-complexity result is that
adiabatic quantum computation can search an unsorted list of $N$
items for a single identifiable item in $O(\sqrt{N})$
time,\cite{Roland} thus matching the speedup of Grover's
algorithm.\cite{Grover} This, however, is a na\"\i ve way of
characterizing the performance of adiabatic quantum computation on
unstructured searches because an $n$th-root speedup still means
that the minimum spectral gap in $\hat{H}(t)$ is exponentially
shrinking, and practically one always needs to operate above some
finite temperature limit.

It is therefore much more realistic when analyzing unstructured
problems to abandon the idea that the system can always remain
overwhelmingly in the instantaneous ground state of $\hat{H}(t)$
and instead view adiabatic quantum computing as an enhanced form
of annealing. The potential for enhancement intuitively derives
from two facts. One, quantum systems can tunnel through energy
barriers instead of waiting for thermal excitation over them. Two,
with a time dependent Hamiltonian $\hat{H}(t)$, one need not face
the same barriers as those in energy landscape $\hat{H}_p$ of
one's problem.

For example, consider an Ising model of the form
\begin{equation}
\hat{H}_p =
\sum_{i,j}J_{ij}\hat{\sigma}_z^{(i)}\hat{\sigma}_z^{(j)}
\label{Ising}
\end{equation}
where the couplings $\{J_{ij}\}$ are randomly drawn with a
zero-mean Gaussian distribution from the interval $[-1,1]$.  If
the Ising model cannot be drawn as a planar graph (with spins as
vertices and nonzero couplings as edges), then finding the ground
state of $\hat{H}_p$ is an NP-hard problem.\cite{Barahona} Quite
general considerations of the distribution of low-lying
excitations of any frustrated system such as that of
Equ.~(\ref{Ising}) imply that simply cooling the system from $T =
\infty$ to $T = 0$ in a time $\tau$ (physically, ``$\infty$''
means ``sufficiently high to make all possible states essentially
equiprobable'') leaves an average residual energy
\begin{equation}
\langle E(\tau) \rangle - E_{ground} \sim \ln(\tau)^{-\xi} \quad
(\xi \leq 2) \label{Class}
\end{equation} in the limit of large $\tau$.\cite{Huse}  Monte Carlo
simulations imply that $\xi$ can be as low as 1 for cooling to the
ground states of some NP-hard random-field Ising
models.\cite{Grest} In contrast, if instead one applies a
transverse magnetic field to the system that decreases from $H =
\infty$ to $H = 0$ in time $\tau$ (all done at $T = 0$), then
according to a model\cite{Santoro} that calculates the residual
energy as the cumulative result of Landau-Zener type
transitions\cite{Landau} at the avoided level crossings between
the instantaneous ground state and first excited states of the
resulting time-dependent Hamiltonian
\begin{equation}
\hat{H}(t) = \hat{H}_p + \Gamma(t) \sum_i\hat{\sigma}_x^{(i)}
\quad (\Gamma(0) = \infty,\ \Gamma(\tau) = 0)
\end{equation}
one finds much improved performance
\begin{equation}
 \langle E(\tau) \rangle - E_{ground} \sim
\ln(\tau)^{-\xi_{QU}} \quad (\xi_{QU} \approx 6) \label{Quant}
\end{equation}The greater size of the exponent $\xi_{QU}$ in
Equ.~(\ref{Quant}) over the exponent $\xi$ in Equ.~(\ref{Class})
is the basis the second major time-complexity result---adiabatic
quantum computation on a given unstructured NP problem should at
least provide cube-root and perhaps as much as sixth-root speedups
over a Metropolis algorithm (simple cooling) approach to the same
problem.

The third, and perhaps most interesting, time-complexity
result---the possibility of exponential speedup on almost all
difficult instances of NP-hard problems---comes from explicit
numerical integrations of Schr\"odinger's equation to simulate
adiabatic computation on sets of small instances of these problems
($\leq 26$ bits is roughly the limit of current supercomputers).
Since the class NP is a worst-case complexity measure, it is
important to consider problem instances that are most likely to be
truly difficult.   For example, Hogg\cite{Hogg} studied random
3-SAT, which is the problem of finding a string of $n$ bits that
completely satisfies $m$ Boolean clauses, each of the form $a$ OR
$b$ OR $c$ where
\begin{enumerate}
\item the literals $a$, $b$, and $c$ refer to 3 distinct bits
chosen uniformly at random

\item each literal, with 50\% probability, contains a negation.
\end{enumerate}
In the limit of large $n$, random 3-SAT instances exhibit a
first-order phase transition from being highly likely to have a
satisfying assignment when $m/n < 4.25$ to being highly unlikely
to have a satisfying assignment when $m/n > 4.25$, and it has been
observed that the most difficult instances to solve appear to fall
on the phase boundary.\cite{Monasson} Hogg studied instances
generated as near as possible to this phase boundary $m/n = 4.25$
and found that for $n \leq 26$ bits (the largest number that could
be numerically analyzed) the median run time scaled only as
$O(n^3)$.

Random 3-SAT instances at the phase boundary still tend to have
exponentially many satisfying solutions which, although they
constitute an exponentially small fraction of the $2^n$ possible
bit strings, one might worry constitute enough degeneracy to
distort study of such small instances of $\leq 26$ bits.  Hence,
the earlier studies of Farhi {\it et al}.\cite{Farhi} and Childs
{\it et al}.\cite{ChildsQIC} focused on randomly generating
NP-complete problem instances with unique satisfying assignments
(``USA instances'') rather than some ratio of constraints to bits.
Median run times scaling only as $O(n^2)$ were observed for both
USA instances of random EXACT COVER up to 20 bits\cite{Farhi} and
random MAX CLIQUE up to 18 bits.\cite{ChildsQIC} (Random EXACT
COVER is the problem of finding a string of $n$ bits that
satisfies $m$ clauses of the form $a + b + c = 1$ where the
literals $a$, $b$, $c$ are distinct bits chosen uniformly at
random.  Random MAX CLIQUE is the problem of finding the largest
subgraph for which every pair of points is connected by an edge
within an $n$-vertex graph constructed such that each pair of
vertices is connected by an edge with 50\% probability.)

The above MAX CLIQUE result may be seen to carry extra
significance because it is widely conjectured there does not exist
any classical polynomial-time algorithm that can identify cliques
in such $n$-vertex random graphs of size $>(1+\epsilon)\log_2(n)$
for any $\epsilon > 0$. In contrast, the maximum clique of an
$n$-vertex random graph is almost always $>\log_2(n)$ and on
average is $2\log_2(n)$. This discrepancy is in fact used as the
basis for a cryptographic scheme.\cite{Juels}

\section{Literature Review: Robustness of Adiabatic Quantum Computation versus Decoherence}
Beyond the potential for significant speedups over classical
computing, which is an attraction that adiabatic quantum computing
shares with other proposed methods of quantum computing, adiabatic
quantum computing is especially attractive for being intrinsically
robust against environmental noise.\cite{ChildsPRA} Firstly, since
an adiabatic algorithm aims to keep its qubits in the
instantaneous ground state of the computational Hamiltonian
$\hat{H}(t)$, dissipation is not intrinsically harmful, but
potentially helpful. (Of course, one cannot depend wholly on
dissipation because then classical thermal annealing could have
solved the problem with equal efficiency.) Secondly, if indeed the
spectral gap of $\hat{H}(t)$ is large enough so that the system
can be kept overwhelmingly in just one energy eigenstate (namely,
the ground state), then the dephasing problem can be essentially
circumvented.   This is because so long as $\hat{H}(t)$ dominates
over the environmental couplings to the qubits, the energy
eigenstates of $\hat{H}(t)$ will be the preferred basis for
dephasing, meaning that an adiabatic quantum computer based on
$\hat{H}(t)$ would never present the environment with a
superposition of states that it could dephase.

Such dominance is easy to achieve in the canonical spin-boson
model with an Ohmic environment having an exponential cutoff after
a frequency $\omega_c$.\cite{Leggett} That is, if $\Delta_o$
denotes the bare energy splitting in the qubits due to coherent
tunnelling between their limiting, classical states in the absence
of coupling to the environment and $\alpha$ represents the
(dimensionless) slope of the environmental spectral density at
$\omega = 0$, then one can conclude non-perturbatively that the
environmentally corrected tunnel splitting $\Delta$ is
\begin{eqnarray}
\Delta &=& \Delta_o \left({\Delta_o \over
\omega_c}\right)^{\alpha/(1-\alpha)} \nonumber \\
&=& \Delta_o\left[1 - \alpha\ln\left({\omega_c \over
\Delta_o}\right) + O(\alpha^2)\right]
\end{eqnarray}
Hence, if $\alpha \ll 1$ and $\omega_c$ is only a few orders of
magnitude above $\Delta_o$, then $\Delta \approx \Delta_o$ meaning
that the qubits' Hamiltonian does indeed dominate over the
environmental couplings and its energy eigenbasis is the preferred
basis for dephasing.

This desirable situation should not be changed by the addition of
$1/f$ or other low-frequency-dominated noise to the Ohmic
environment. Components in the environmental spectral density that
are of much lower frequency than the bare tunnel splitting
$\Delta_o$ denote environmental dynamics that are much slower than
that of the qubits.  Therefore, essentially regardless of the
qubit-environment coupling, these components can only become
entangled to time-averaged properties of the qubits such as their
energy.\cite{Paz}  While low-frequency components could thus
significantly change dephasing rates, they should not change the
fact that the environment respects the energy eigenbasis of the
qubits' Hamiltonian as the preferred basis for dephasing, meaning
that adiabatic algorithms would still essentially circumvent the
dephasing problem.

\section{Main Result: A Scalable Architecture for Adiabatic Quantum Computing}

Another special advantage of adiabatic quantum computing is the
main point of this paper: there exists a simple, scalable
architecture for adiabatic quantum computation that can handle
NP-hard problems without requiring local coherent manipulations of
the qubits.  The starting point for the architecture is the fact
that it is NP-hard to calculate the ground state of a 2D
antiferromagnetically coupled Ising model placed in a magnetic
field even if each Ising spin is coupled to no more than 3
others.\cite{Barahona} In other words, if $G = (V,E)$ is a planar
graph with vertices $V$ and edges $E$ and even if every vertex in
$V$ belongs to $\leq 3$ edges in $E$, then it is NP-hard to find
the ground state of
\begin{equation}
\hat{H} = \sum_{i \in V}\hat{\sigma}^{(i)}_{z} + \sum_{(i,j) \in
E} \hat{\sigma}^{(i)}_z \hat{\sigma}^{(j)}_z
\end{equation}
which is an Ising model with spins at the vertices $V$, equal
strength antiferromagnetic couplings along the edges $E$, and a
homogeneously applied field giving each spin a Zeeman splitting
equal to the antiferromagnetic coupling strength.  (The direction
of this applied field sets the ``$z$-axis.'')

This proof of NP-hardness isomorphically maps the above Ising
model onto the MAX INDEPENDENT SET problem, which is the problem
of finding for a graph $G = (V,E)$ the largest subset $S$ of the
vertices $V$ such that no two members of $S$ are joined by an edge
from $E$. Solving MAX INDEPENDENT SET for a given graph $G =
(V,E)$ is completely equivalent to solving MAX CLIQUE for the
graph's complement $G^c \equiv (V,(V \times V)/E)$. (In other
words, $G^c$ connects with edges all the pairs of vertices that
were unconnected in $G$ and disconnects all the pairs were
connected.) Therefore, the MAX CLIQUE
results\cite{ChildsQIC}$^,$\cite{Juels} referenced above imply
that MAX INDEPENDENT SET is an equally interesting problem for
treatment by adiabatic quantum computation, likely admitting not
even a rough approximation in polynomial time classically, yet
apparently admitting efficient solution by quantum adiabatic
evolution---at least in small instances (and thus, hopefully, for
almost all, if not all, larger instances).

There exists a simple, regular architecture that can be programmed
to pose the above NP-hard Ising model ground state problem for any
instance with $\leq n$ spins.  One way to see this is to recall
that any planar graph $G$ with $\leq 3$ edges per vertex admits a
topologically equivalent embedding $Q$ in a square grid of $\leq
9n^2$ vertices.\cite{Valiant}  In other words, there exists a map
$Q$ that $(1)$ maps each vertex $v$ of $G$ onto a distinct point
$Q(v)$ of the grid and $(2)$ inserts ``dummy vertices'' so as to
map an edge $\bf{e}$ connecting vertices $u$ and $v$ of $G$ into a
path $Q(\bf{e})$ on the square grid such that $Q(u)$ and $Q(v)$
are its endpoints and such that if $\bf{e}$ and $\bf{f}$ are two
distinct edges of $G$, then $Q(\bf{e})$ and $Q(\bf{f})$ do not
intersect (except, obviously, if $\bf{e}$ and  $\bf{f}$ share an
endpoint, in which case $Q(\bf{e})$ and $Q(\bf{f})$ will also
share an endpoint).

\begin{figure}[b]
\includegraphics*[0.5in,0.5in][4.5in,4in]{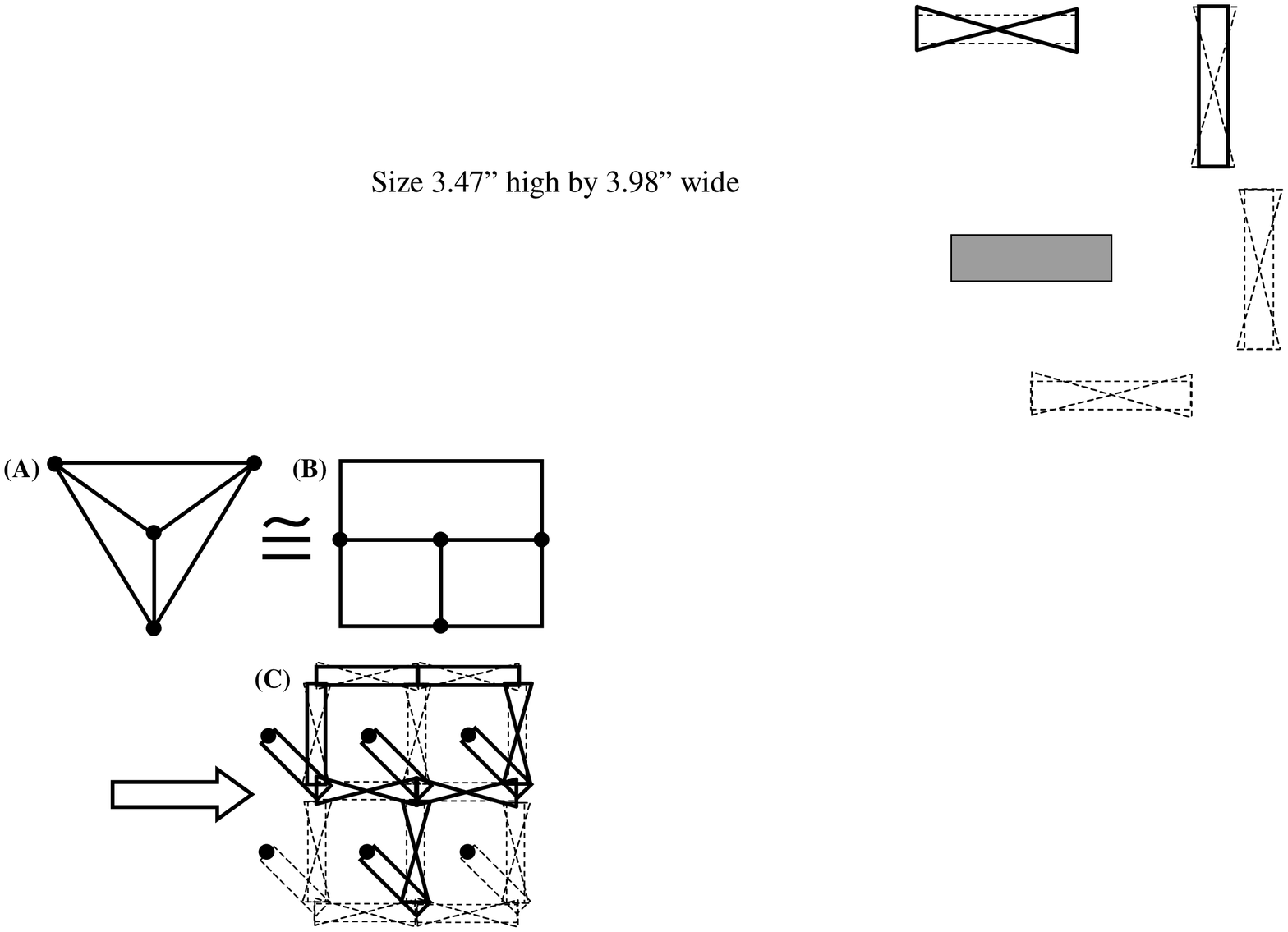}
\caption{\bf{(A)} A regular 4-vertex graph for an Ising model.
Vertices denote location of qubits, and edges denote
antiferromagnetic couplings. \bf{(B)} Topologically equivalent
square grid embedding. (C) Possible implementation. Boldface lines
denote couplings switched on, dashed lines denote couplings
switched off.}
\end{figure}

Thus, one simple, regular architecture would be of the type
depicted below in Fig. 1: two stacked square grids of
ferromagnetic and antiferromagnetic couplings, respectively,
having at the center of each square a qubit with a ferromagnetic
coupling to, say, the inductive couplings meeting at lower right
corner of the square. Since a useful adiabatic scheme would be one
like Equ.~(3) which merely adds a homogeneous, steadily decreasing
transverse magnetic field to the problem Hamiltonian, the coherent
evolution of the qubits could be affected simply with a globally
applied magnetic field.  This avoids the need both for any local
coherent manipulations as well as for \textit{tunable} rather than
merely \textit{switchable} couplings.  Furthermore, as the the
switching of the couplings is necessary only for programming the
desired instance to be solved, it can be done ``offline,''
\textit{before} any coherent manipulation of the qubits begins,
thus obviating the need for ``quiet'' switches.

A final consideration for scalability is to calculate how much
imprecision can exist in the parameters of the qubits, the
couplings, and the applied field before one loses the ability to
encode reliably in the above architecture any desired $n$-bit
instance of MAX INDEPENDENT SET.  To answer this, first recall
that Ref. 5 proved that the maximum independent sets of a graph $G
= (V,E)$ are in 1-to-1 correspondence with the ground state(s) of
the Hamiltonian in Equ.~(6), \textit{i.e.}, an Ising model with
spins at the vertices $V$, antiferromagnetic couplings all of
equal strength along the edges $E$, and a homogeneous applied
magnetic field giving every spin a Zeeman splitting equal to the
antiferromagnetic coupling strength. However, so long as one is
satisfied with obtaining any one of the maximum independent sets
of $G$ (which is reasonable since obtaining any one is NP-hard),
one does not need to calibrate so precisely the couplings among
themselves and to the applied field because the ground state(s) of
any Ising model from the family
\begin{equation}
\hat{H} = \sum_{i \in V} J_i \hat{\sigma}^{(i)}_z + \sum_{(i,j)
\in E} J_i J_j \hat{\sigma}^{(i)}_z \hat{\sigma}^{(j)}_z
\end{equation}
where $\{J_i\}$ are \textit{arbitrary} positive scalars will also
be ground state(s) of the Equ.~(6) Ising model and hence maximum
independent set(s) of $G$. Hence, in regard to programming the
computer so that its final ground state has a large overlap with
at least one solution of a desired instance of MAX INDEPENDENT
SET, the proposed architecture is robust against
\textit{arbitrary} variance in the magnetic moments of the qubits.
(Note, however, that while variance in $\{J_i\}$ poses no problem
in programming the computer, it might still cause a problem in
ensuring that the computer produces accurate solutions quickly.
Specifically, as the variance becomes larger, the problem
Hamiltonian (7) looks more and more like random field magnet than
a ``structured'' NP problem, likely lessening the chances that the
adiabatic scheme $\hat{H}(t)$ will keep a large spectral gap and
thus permit a running time exponentially faster than any classical
algorithm.)

More general errors can come from inhomogeneities in both the
magnitude and the direction of the applied magnetic field and from
inductive couplings beyond the desired ones between nearest
neighbors. Thus, to Equ.~(7) one can add the general perturbation
\begin{equation}
\hat{H}' = \sum_{k \in \{x,y,z\}} \sum_{i \in V} \epsilon^{(i)}_k
\hat{\sigma}^{(i)}_k + \sum_{k,l \in \{x,y,z\}} \sum_{i
> j} \delta_{kl}^{(ij)} \hat{\sigma}^{(i)}_k \hat{\sigma}^{(j)}_l
\end{equation}
Note that Pauli operator terms such as $\hat{\sigma}^{(i)}_k$ or
$\hat{\sigma}^{(i)}_k \hat{\sigma}^{(j)}_l$ merely prescribe bit
and/or phase flips in the eigenbasis of the Equ.~(7) problem
Hamiltonian, meaning that each of the perturbing terms in
$\hat{H}'$ couples an unperturbed ground state to only one other
unperturbed eigenstate. Therefore, if $n$ denotes the number of
qubits, $\lambda$ denotes the largest of the scalars
$\{\epsilon^{(i)}_k, \delta_{kl}^{(ij)}\}$ in Equ.~(8), and
$\Delta$ denotes the minimum increase in energy caused by flipping
1 or 2 qubits of the ground state (thus, $\Delta \approx \min_i
[2J_i$]), then $\lambda n^2 / \Delta$ serves as a simple upper
bound on the magnitude of the first order correction to the ground
state wavefunction.  Higher order corrections should be
well-behaved since performing 1 or 2 bit flips on any excited
state also typically entails an energy change of at least
$\Delta$, implying that the entire perturbation series correction
should essentially converge like the geometric series $\sum_k
(\lambda n^2 / \Delta)^k$. Hence, the architecture's tolerance for
general imprecision in the qubit couplings and inhomogeneity in
the applied field decreases quadratically in the number of qubits.
(If unwanted inductive couplings can be limited to a constant
number of ``next-nearest'' neighbors, then the tolerance shall
decrease only linearly.)

\section{Conclusion}

We have reviewed the theory of adiabatic quantum computation
regarding its potential speedup over classical computation and its
greater robustness toward noise over other forms of quantum
computation.   We then proposed a scalable architecture for
adiabatic quantum computation based on the NP-hardness of
calculating the ground state of a planar antiferromagnetically
coupled Ising model placed in a magnetic field.  Clearly, such an
architecture could be used with any qubit whose states have
opposite magnetic dipole moments.  Future work will present a
detailed proposal for its implementation using superconducting
persistent-current qubits,\cite{Mooij} which constitute a current,
promising approach to lithographable solid-state qubits.

\begin{acknowledgments}
This work is supported in part by ARDA and DoD under the AFOSR
DURINT Program. WMK gratefully acknowledges fellowship support
from the Fannie and John Hertz Foundation.
\end{acknowledgments}


\end{document}